\documentclass[conference]{IEEEtran}
\IEEEoverridecommandlockouts

\usepackage{cite}
\usepackage{graphicx}
\usepackage{amsmath,amssymb,amsfonts}
\usepackage{booktabs}
\usepackage{textcomp}
\usepackage{xcolor}
\usepackage{url}
\usepackage{comment}

\def\BibTeX{{\rm B\kern-.05em{\sc i\kern-.025em b}\kern-.08em
    T\kern-.1667em\lower.7ex\hbox{E}\kern-.125emX}}

\title{Scaling DeFi with ZK Rollups:\\Design, Deployment, and Evaluation of\\a Real-Time Proof-of-Concept}

\author{
\IEEEauthorblockN{
Krzysztof M. Gogol, Szczepan Gurgul, Faizan Nehal Siddiqui, David Branes, Claudio J. Tessone}
\IEEEauthorblockA{University of Zurich}
}

\begin{document}

\maketitle

\begin{abstract}
Ethereum's scalability limitations pose significant challenges for the adoption of decentralized applications (dApps). Zero-Knowledge Rollups (ZK Rollups) present a promising solution, bundling transactions off-chain and submitting validity proofs on-chain to enhance throughput and efficiency. In this work, we examine the technical underpinnings of ZK Rollups and stress test their performance in real-world applications in decentralized finance (DeFi). We set up a proof-of-concept (PoC) consisting of ZK rollup and decentralized exchange, and implement load balancer generating token swaps. Our results show that the rollup can process up to 71 swap transactions per second, compared to 12 general transaction by Ethereum. We further analyze transaction finality trade-offs with related security concerns, and discuss the future directions for integrating ZK Rollups into Ethereum's broader ecosystem.
\end{abstract}


\section{Introduction}
\label{sec:introduction}

Ethereum, launched in 2015 by Vitalik Buterin~\cite{Buterin2014Ethereum}, was the first programmable blockchain, marking a significant evolution from Bitcoin~\cite{Nakamoto2008Bitcoin}, the first blockchain, primarily designed for peer-to-peer payments. Ethereum introduced the concept of \textit{smart contracts}~-~self-executing computer programs that run on the blockchain~\cite{Gogol2023SoKDeFi}~-~paving the way for decentralized applications (dApps). These dApps provide a transparent, immutable environment for complex transaction processing without intermediaries. Over the years, Ethereum has become the backbone of the decentralized finance (DeFi)~\cite{werner2022sok,schaer2023defimarkets,Auer2023technologydefi} with more than 100 billion USD in assets locked in various DeFi protocols~\cite{2024DeFiLlama}.

However, Ethereum's growing popularity has also led to significant challenges. Ethereum's limited throughput and high transaction costs, particularly during periods of network congestion, hinder the user experience and pose significant barriers to mass adoption~\cite{Bez2019Scalability}. To address these limitations, \textit{Layer-2 (L2) scaling solutions} have emerged as a viable method for improving transaction scalability while preserving the security and decentralization.
Among L2 solutions, \textit{rollups}~-~which perform transactions off-chain and then record transaction data on Ethereum~-~have become the dominant approach. Rollups are categorized into two types: \textit{optimistic rollups} and \textit{ZK rollups} based on the transaction validity mechanism they apply. 

This study focuses on ZK rollups, which utilize zero-knowledge (ZK) proofs to verify transactions, offering enhanced scalability and security compared to their optimistic counterparts~\cite{thibault2022blockchain}. 
ZK rollups aggregate multiple transactions off-chain and submit a validity proof to Ethereum. This ensures that all transactions adhere to network rules without requiring individual on-chain execution, thereby reducing gas costs and increasing transaction throughput. These improvements make ZK Rollups essential for addressing Ethereum’s scalability needs while maintaining its decentralized and fast transaction finality~\cite{yee2022shades}. Several implementations of ZK rollups exist, including \textit{ZKsync} by Matter Labs~\cite{zkSyncEra}, \textit{zkEVM} by Polygon, and \textit{StarkNet} by StarkWare. Among EVM-compatible ZK Rollups, ZKsync leads in DeFi total value locked (TVL)~\cite{2024DeFiLlama}.

Rollups overall already attracted half of TVL that is currently locked in Ethereum (ca \$70bn)~\cite{2024DeFiLlama}, and the average daily transaction per second (TPS) of rollups is 30 times higher than Ethereum~\cite{2024L2BeatTPS}. These metrics however, do not tell us much  about the complexity of the underlying transaction. Thus, novel metrics~-~\textit{gas per second} (GPS)~-~was introduced to measure the actual processing capabilities of blockchains. The average GPS of rollups is 55 times higher than Ethereum, on average~\cite{2024RollupGPS}. In this research, we perform the first stress tests of rollup using swap transaction at decentralized exchange (DEX)~\cite{Xu_2023}, instead of simile token transfers between wallets to measure the maximum GPS that ZK rollups can achieve. 


\subsection{Related Work}
This paper focuses specifically on a single category of L2 solutions based on ZK proofs. For a comprehensive introduction to various L2 architectures, we refer readers to Thibault~\cite{thibault2022blockchain}. For a detailed exploration of optimistic rollups, Armstrong~\cite{armstrong2021} offers valuable insights.

Chaliasos~\cite{Chaliasos2024} offer a comprehensive analysis of the security implications and performance bottlenecks of zk-rollups, with a focus on transaction finality and data availability issues. Their study highlights the practical challenges and technological advancements required to ensure robust security and data integrity in zk-rollup implementations. This work is essential for understanding the broader limitations and potential improvements necessary to enhance the scalability and resilience of L2 solutions.

Parallel to this, Gogol~\cite{Gogol2024AMM} investigates arbitrage dynamics within Automated Market Makers (AMMs) on zk-rollups, contrasting the operational frameworks of centralized exchanges (CEXs) and decentralized exchanges (DEXs). This research is particularly relevant to our study, as it provides insights into the economic behaviors and transaction efficiency on L2 solutions, which are critical for understanding the impact of zk-rollups on market liquidity and trading strategies.

\subsection{Contribution}
This paper provides an analysis of ZK rollups, exploring their architecture, evaluating their real-time performance, and discussing applications in DeFi, and beyond. We set up a ZK rollup utilizing the ZKsync framework and forked Uniswap, a leading decentralized exchange (DEX) in (DeFi)~\cite{xu2021sok}. Subsequently, we created liquidity pools comprising ERC-20 tokens and implemented a smart contract with a load balancer designed to autonomously generate swap transactions. 
\begin{itemize}
    \item With this setup, we empirically measured the efficiency and security of the rollup. Our results show that ZK rollups can process up to 71 swap transactions per second, compared to 12 general transaction by Ethereum. 
    \item We examined trade-offs between scalability, decentralization, and security.
    \item Finally, all code and modifications done to the Uniswap code are made publicly available, allowing the reproductability of research.
\end{itemize}

\subsection{Paper Organization}
The structure of this paper is organized as follows. Section~\ref{sec:background} introduces the essential information about rollups, Section~\ref{sec:system} presents the test architecture for a proof-of-concept, Section~\ref{sec:performance_simulation} simulated transactions and Section~\ref{sec:stress-test} - stress-testing outcome. Subsequently, Section~\ref{sec:discussion} examines the results, and Section~\ref{sec:conclusions} offers the conclusions.
\section{Background: ZK Rollup}
    \label{sec:background}

A blockchain is a decentralized network that maintains a shared, immutable database. This system allows only the addition of new transactions, disallowing any alterations or deletions of existing ones. Transactions are organized into blocks, and the sequence of these blocks is preserved and verified using cryptographic hash functions and digital signatures. There are two general approaches for scaling blockchain solution. Layer-1 (L1) blockchain scaling involves development of new blockchain, often new consensus mechanism~\cite{Lashkari@2021Consensus}, sharding~\cite{Wang@2019Sharding} and its own physical infrastructure. Layer-2 (L2) blockchain scaling approach leverages the decentralization and security of existing L1 blockchains, while conducting complex calculations outside of L1 network~\cite{sguanci2021layer2,gangwal2022survey}.

\paragraph{Layer-2 Blockchain} 
Layer-2 (L2) blockchains are scaling solutions built on top of core L1 blockchains, such as Ethereum. L2 solutions are intended to preserve the security and decentralisation of the underlying network while addressing the intrinsic drawbacks of L1 blockchains, like high transaction fees, constrained throughput, and latency. L2 solutions group several transactions together, offload transaction processing off the L1 blockchain, and settle the outcomes on the L1 chain on a regular basis. 
There are various ways by which this is achieved, notable techniques are optimistic rollups, zk-rollups, side-channels and plasma. Each of these techniques utilize its own logical process to execute transactions off-chain, batch them, and settle them on L1 in a secure and efficient way. As a results, traffic on the main chain is drastically reduced and far more transactions can be processed~\cite{sguanci2021layer2,gangwal2022survey}.

\paragraph{Zero-Knowledge Proof} 

Zero-Knowledge Proofs (ZKPs) is a cryptographic system that enables one party (the prover) to demonstrate to another (the verifier) that a certain assertion is true without disclosing any further information beyond the statement's veracity. The prover has a piece of secret information (e.g., a private key or solution to a computation) and wants to prove they possess it and the verifier checks the proof provided by the prover to confirm the claim's validity without learning the secret information.


\paragraph{zk-SNARKs} 

zk-SNARKs (Zero-Knowledge Succinct Non-Interactive Arguments of Knowledge) are a subset of zero-knowledge proofs, their unique properties (succinctness and non-interactivity) make them particularly suited for blockchain and decentralized systems. They enable highly efficient and scalable applications without compromising on privacy. 
The \texttt{Succinct} part makes sure that verification is always done quickly as compared to computation. The second important feature is \texttt{Non-Interactive}, which limit the back and forth interaction between the prover and verifier. Interactive proving requires both parties to have interaction, which limits its power. zk-SNARKs are widely used in privacy-focused and scalable blockchain technologies, such as ZK rollups and privacy coins like Zcash.

\paragraph{ZK rollup} 

\begin{figure}[h]
\centering
  \includegraphics[width=0.8\columnwidth]{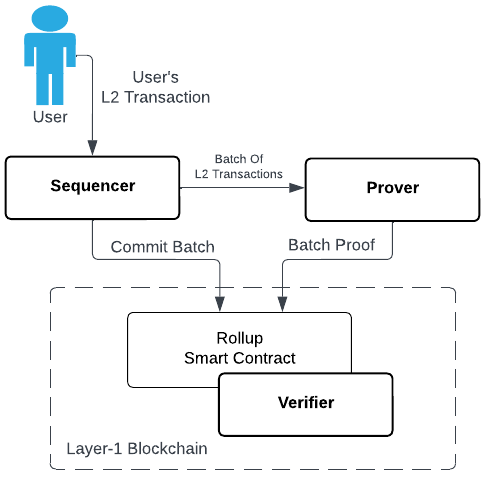}
    \caption{High-level architecture of ZK Rollups. A sequencer bundles and executes transactions off-chain, while the prover generates cryptographic proofs for on-chain verification. This structure enables scalable and trust-minimized Layer 2 computation.
  \cite{formalfoundation}}
  \label{fig:highlevelarch}
\end{figure}

ZK rollup~\cite{Thibault@2022rollups} is a L2 scaling solution that batches transactions together on L2 and then utilizes zero-knowledge proofs(ZKPs) to verify the batch on L1. It generates a zero-knowledge proof for each batch, which verifies the validity of all transactions in the batch. This proof is sent to a smart contract on L1 which validates the proof. 
ZK rollup uses zk-SNARKs as ZKP, as succinct proofs are small and quick to verify which makes them more efficient. zk-SNARKs are also non-interaction so the prover and verifier don’t need to communicate back and forth.
The high-level architecture of ZK rollup is presented in Figure~\ref{fig:highlevelarch} with the specially deployed sequencers and prover. An externally owned address (EOA) is a type of account that is controlled by a private key and is used by individuals to send transactions, keep an accounting of their account, and interact with smart contracts. 

As presented in Figure~\ref{fig:highlevelarch} ZK rollup is composed of L2 components like sequencers and proves, where the sequencer is responsible for processing and bundling transactions into batches that are then submitted to the L1 blockchain~\cite{formalfoundation}. The prover generates cryptographic proofs that prove the correctness of these batches. On L1, there are smart contracts deployed that serve as a foundation layer for managing and executing the ZK rollup. The L1 smart contracts presented in figure~\ref{fig:highlevelarch} ensure that the batches are recorded on L1 for accountability. If the proof received is valid, the L1 state would be updated accordingly. Forcing transactions by EOA allows direct submission of a transaction to the L1 smart contract if the sequencer is unavailable or unresponsive.

\paragraph{Sequencer} 

Sequencers~\cite{motepalli2023soksequencers} are the most important component in a L2 blockchains as they are responsible for ordering the transactions on L2, batching them together, executing them on L2 and finally submitting them to L1. They also ensure that these transactions are included in the blockchains in a timely and efficient manner. In essence, a sequencer is responsible for ensuring that transactions on L2 are processed efficiently, maintaining the integrity and speed of the L2 network.
After batching and executing the transactions they are also responsible for updating the L2 state. This includes keeping track of account balances, smart contract executions, and other state changes resulting from transactions. Unlike L1, sequencers don’t get congested by transactions and don’t face high gas like L1.

\paragraph{Prover} 

Prover is one of the two core components in ZK rollups. Its purpose is to generate cryptographic proofs for the transactions that have been processed off-chain, which is then used for verifying those transactions on L1. In case for ZKsync, the prover uses zero-knowledge proofs which is known as zk-SNARK
The sequencer in l2 batches the transactions together and then processes them which significantly reduces the processing cost. Prover then works by generating a proof for each of these batches using zk-SNARKs. This generated proof is submitted to the L1 blockchain with a transaction, which is then verified by a smart contract on L1. 
Prover helps in ensuring that L2 blockchain is trustworthy and secure along with it being scalable. Moreover, it also ensures privacy as the proofs can be verified without revealing any detail.
\section{System Architecture and Design}
    \label{sec:system}

In order to evaluate the ZK rollup applicability to DeFi we set up a ZK rollup, deploy there test ERC-20 tokens and a DeFi protocol - a decentralized exchange (DEX) that allows to swap our test ERC-20 tokens for each other. Next, we stress the system with automatically generated swaps.

\subsection{ZKsync As ZK Rollup}

There are various provides of ZK rollups - StarkNet, Polygon zkEVM, ZKsync, but most of the ZK rollups are not compatible with Ethereum Virtual Machine (EVM)~\cite{2024L2BeatTPS}. As vast majority DeFi protocols are developed in Solidity for EVM and DeFi total value locked (TVL) resides in majority on EVM-chains~\cite{2024DeFiLlama}, we decided to set-up the EVM zk-rollup: ZKsync, which currently has the highest TVL in DeFi among zk roll-ups~\cite{2024DeFiLlama}.

We use the elastic chain framework offered by MatterLabs, one of the contibutors to ZKsync development. It is important to undelrying that we do not fork ZKsync - a public zk-rollup, but set up a new zk rollup (with new genesis block) using the code base of ZKsync. The setup is configured, deployed and running on top of the Ethereum L1 network, in our case Sepolia testnet. The chain operates independently with its own sequencer and prover. 

\subsection{Uniswap As DEX}
In this setup, we have forked Uniswap V2, a widely-used decentralized exchange protocol, onto one of our ZKsync chain. This involves replicating the Uniswap V2 smart contracts and deploying them on our L2 to facilitate automated token trading within our L2. 

As ZK-sync is EVM-compatible, but not EVM-equivalent chains, the successful deployment of Uniswap V2 on ZKsync required adjustments to the original codebase, including Solidity version upgrades and integration adjustments. These changes ensured compatibility with ZKsync while preserving the core functionality of Uniswap. The deployment process, from setting up the contracts to creating pairs and performing swaps, highlights the adaptability of Uniswap V2 for L2 environments.

\section{Experimental Setup}
\label{sec:performance_simulation}
This section outlines the experimental framework used to evaluate the performance of our ZK Rollup-based proof-of-concept (PoC). We describe the infrastructure configuration, transaction generation process, and data extraction pipeline supporting the benchmarking of real-time decentralized exchanges.

\subsection{Infrastructure Specifications}

Our ZKsync-based prototype was deployed on a high-performance GPU server tailored to meet the computational demands of zk-rollup processing. The machine was equipped with 16 CPU cores, 64 GB of RAM, a high-performance virtual machine instance (HPCv3), an NVIDIA T4 GPU, and 300 GB of HDD storage. This configuration ensures the system can handle the intensive tasks associated with zero-knowledge proof generation, transaction batching, and data availability. This setup was selected based on the technical requirements of ZKsync for optimal performance, ensuring sufficient computational power and storage to handle the demanding tasks of zk-rollup processing, transaction batching, and cryptographic proof generation. The ZKsync nodes were installed and configured to support the operation of our L2 chains as described below.

\subsection{Transaction Generation Methodology}

The primary goal of the stress-testing framework is to benchmark the maximum achievable TPS on our custom ZKsync deployment integrated with a Uniswap V2 fork. To provide a more realistic performance evaluation, we use swap transactions—which are computationally and gas-intensive—rather than simple token transfers. The implementation is based on the public generator code by Bogatyy et al.~\cite{bogatyy2024generator}, with modifications available in our repository~\cite{transactionGenerator}.

Our benchmark simulates 1-hop token swaps using the \texttt{swapExactTokensForTokens} function from the UniswapV2Router. Swap execution is synchronized by introducing a delay, ensuring parallel transaction submission. Transactions are sent from multiple IP addresses, configured via \texttt{iptables}, to emulate distributed real-world traffic and avoid overloading a single RPC endpoint. Communication is handled via a WebSocket connection to the local ZKsync node (port 3051), chosen for its lower latency and reduced overhead compared to HTTP—both critical for high-throughput scenarios.

To avoid overloading the system, we limited the number of generator instances to five, with each instance tied to a separate IP address. We observed that increasing beyond five instances led to sequencer instability, sometimes even causing it to fail. This confirms the sequencer as a potential single point of failure in current rollup designs. As a result, five instances were used consistently in our evaluation.

Table~\ref{tab:transaction-generation} presents the performance results across different instance counts. While throughput initially scales with more instances, performance plateaus and eventually declines as system overhead grows, revealing the sequencer’s bottlenecks under parallel load.

\begin{table}[h]
    \centering
    \caption{Performance of the ZK Rollup Transaction Generator. Peak TPS is achieved with a single instance; throughput declines as sequencer load increases, exposing bottlenecks in parallel processing.}
    \begin{tabular}{lrrr}
        \toprule
        Instances run & Transactions & Time [s] & TPS \\
        \midrule
        1 & 200 & 1 & 200.00 \\
        2 & 400 & 2 & 200.00 \\
        3 & 600 & 7 & 85.71 \\
        4 & 800 & 13 & 61.54 \\
        \textbf{5} & \textbf{1,000} & \textbf{14} & \textbf{71.43} \\
        \bottomrule
    \end{tabular}
    \label{tab:transaction-generation}
\end{table}

To generate the transactions, we created 50 accounts derived from a single mnemonic phrase. Each account was funded with native tokens to cover gas fees. Prior to execution, all 50 addresses pre-approved token spending on the liquidity pool to enable seamless swap execution. Each account performed 20 swaps, totaling 1,000 swap transactions.

All transactions were broadcasted concurrently from the five generator instances. Transaction data, including hashes and sender addresses, was logged during execution for post-analysis. This coordinated submission guarantees that identical swap transactions from multiple accounts are injected simultaneously into the network, creating a realistic high-load scenario for evaluating system throughput and latency.

\subsection{Data Logging and Parsing}

After executing the swap transactions via the stress-testing setup, all relevant data must be parsed from the blockchain to compute throughput (TPS) and analyze transaction latency. During the execution phase, each generator instance (mapped to a distinct IP address) logs transaction data—such as sender addresses and transaction hashes—into separate log files. These logs are then merged into a single consolidated file to enable unified analysis.

To extract on-chain information, a custom blockchain parser processes the log data by issuing batch requests to retrieve transaction metadata, including block timestamps, block numbers, block sizes, and the time of transaction submission. Each transaction hash is mapped to the block in which it was included. The resulting dataset—a list of swap transaction hashes sorted by their send time—contains both the timestamp of submission and the timestamp of inclusion, making it possible to calculate precise latency metrics.

This structured data enables the evaluation of both throughput and inclusion delay over time. To support visualization, the parsed data is loaded into a Pandas DataFrame and grouped into 0.1-second intervals to smooth the signal. Timestamps are normalized to seconds relative to the earliest transaction for temporal alignment. The final plots depict two key lines: transactions sent and transactions included, with time on the x-axis and cumulative transaction count on the y-axis. These visualizations form the basis for performance insights shown in Figure~\ref{fig:transactiongenerator}.

\section{Performance Evaluation}
\label{sec:stress-test}

We evaluate the performance of our ZKsync-based Layer-2 system through a series of stress tests. We extract on-chain swap transaction data to analyze system throughput, latency, and block utilization.

\subsection{Throughput Analysis}

Transactions per second (TPS) is a common metric used to quantify the throughput of blockchain systems. It measures how many transactions a system can successfully process within one second, offering a practical lens through which to assess scalability and performance. In our setup, TPS is calculated cumulatively over time using the following formula:
\[
\text{Cumulative TPS} = \frac{\text{Total (swap) transactions processed}}{\text{Elapsed time (in seconds)}}
\]
Unlike standard TPS benchmarks that often rely on simple token transfers, our measurement is based on swap transactions at DEX—complex, gas-intensive operations that more closely reflect real-world DeFi usage, making our results more indicative of practical throughput under realistic load conditions.

\begin{figure}[h]
\centering
  \includegraphics[width=1\columnwidth]{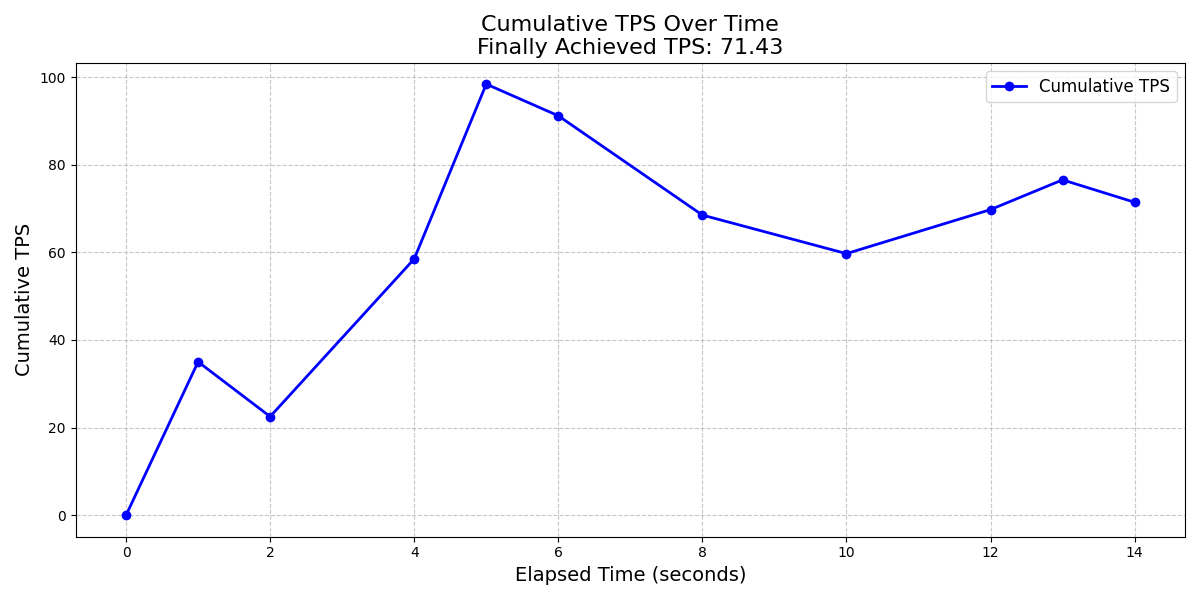}
  \caption{Cumulative (swap) transaction throughput. Peak TPS of 98.4 achieved at 4 seconds, stabilizing at 71.43 TPS by end of test. Demonstrates the experimental system’s sustained throughput under load.}
  \label{fig:transactiongenerator-tps}
\end{figure}

As shown in Figure~\ref{fig:transactiongenerator-tps}, the cumulative TPS of the system steadily increases, peaking at 98.4 TPS around the 4-second mark. Following this initial surge, it stabilizes, fluctuating slightly before settling at 71.43 TPS by the end of the 14-second test window. This final value serves as a reliable estimate of the system’s sustained throughput capacity under consistent load. 

The results demonstrate the ZKsync system’s ability to efficiently process a high volume of gas-intensive transactions over time, validating its applicability to performance-critical DeFi scenarios. The achieved cumulative TPS of 71.43, significantly surpass Ethereum's native estimated throughput of 15 TPS, demonstrating that ZK rollups can effectively scale Ethereum by handling higher transaction volumes while preserving its foundational properties such as security and decentralization.

\subsubsection*{\textbf{Key Finding}: TPS Metrics Vary Widely Between Simple Transfers and Gas-Intensive Swaps}

Unlike conventional throughput benchmarks that rely on lightweight token transfers, our evaluation uses swap transactions on DEX—operations that are significantly more complex and gas-intensive. This difference is critical, as it reflects more realistic DeFi usage and provides a better indicator of how rollup systems perform under genuine load conditions.
For example, ZK rollups such as ZKsync claim theoretical capacities of up to 2,000 TPS, while optimistic rollups like Arbitrum have reported reaching over 40,000 TPS in synthetic benchmarks. However, these figures are often based on ideal conditions using minimal transaction types that do not reflect actual on-chain activity.

In contrast, our stress test floods the system exclusively with swap transactions. As illustrated in Figure~\ref{fig:transactiongenerator-tps}, our experimental system peaks at 98.4 TPS within 4 seconds and stabilizes at 71.43 TPS by the end of the test. This outcome underscores the platform’s ability to handle complex, computation-heavy DeFi workloads, providing a grounded assessment of sustainable throughput.
These findings suggest that to accurately evaluate Layer 2 scaling solutions, benchmarks must account for transaction complexity—not just raw volume.

\subsection{Latency and Finality}

\begin{figure}[h]
\centering
  \includegraphics[width=1\columnwidth]{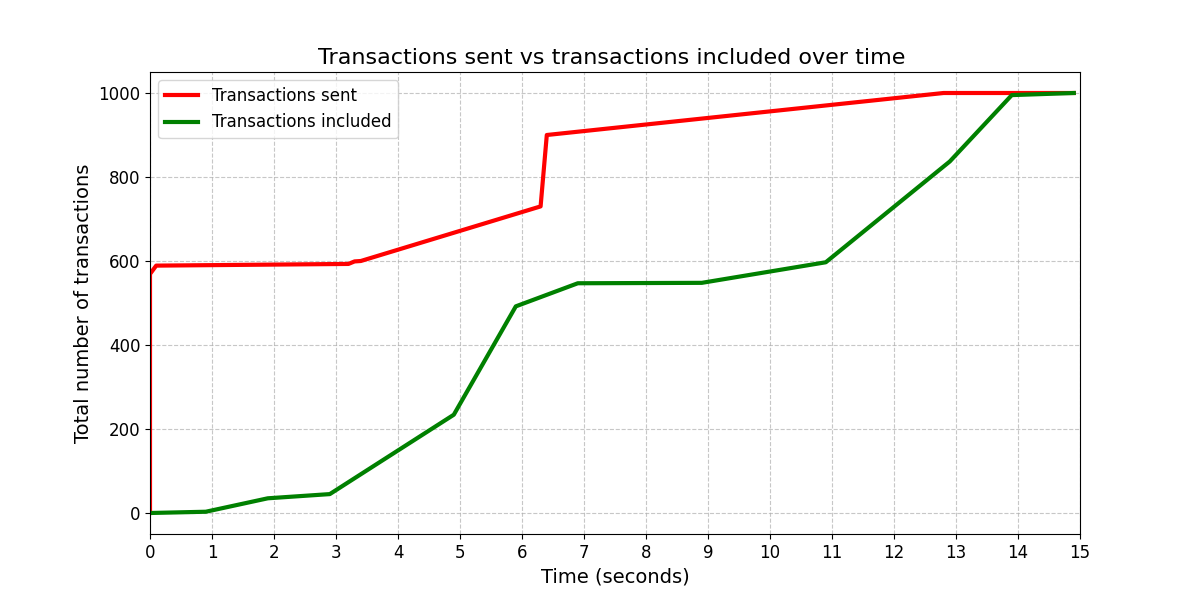}
  \caption{Transactions sent vs. included. Highlights latency during peak load and eventual consistency. The system stabilizes within 15 seconds.}
  \label{fig:transactiongenerator}
\end{figure}

Figure~\ref{fig:transactiongenerator} plots two lines: the red line represents transactions sent by the generator, while the green line shows the inclusion of those transactions into blocks over time. At the start of the test, the generator sends an initial burst of approximately 600 swap transactions almost instantaneously. This sudden influx temporarily overwhelms the sequencer, which pauses new submissions while it processes the backlog and begins forming batches.

Following this initial burst, a second wave of transactions begins around the 3-second mark and continues steadily until all are sent by 13 seconds. The lag between the red and green lines reveals two key dynamics of zk-based rollups: transaction latency and finality.

\subsubsection*{Latency}  
The delay between when a transaction is sent and when it is included in a block reflects the system’s responsiveness under load. In the first few seconds, latency is high due to the sequencer catching up. However, by 7 seconds, the gap between sent and included transactions narrows as the system stabilizes and enters a steady state. Full inclusion is completed by 15 seconds, meaning all transactions sent have been successfully processed, though not yet finalized on Ethereum.

\subsubsection*{Finality.}  
In rollup-based systems like ZKsync, we distinguish between two levels of finality: soft and hard.
\emph{Soft finality} occurs when a transaction is included in an L2 block by the sequencer. At this point, the transaction is visible and assumed to be final under normal operating conditions—but it can still be reversed if the batch is invalidated before being submitted to L1.
\emph{Hard finality} is achieved only when the corresponding batch, along with its validity proof, is successfully posted and verified on Ethereum L1. This ensures that the transaction is cryptographically secured and permanently immutable.

In our test, all transactions reached soft finality within 15 seconds, reflecting the system’s ability to maintain throughput and responsiveness under load. Hard finality, however, depends on the timing of the batch submission to L1, which occurs asynchronously and is not captured directly in this plot.

\subsubsection*{\textbf{Key Finding}: Instant User Experience vs. Delayed Trust Guarantees}
This distinction is crucial: while L2 systems provide near-instant user experience via soft finality, their trust guarantees ultimately rely on the timely and correct execution of batch submissions to Ethereum for hard finality.

\subsection{Block Utilization}

\begin{figure}[h]
\centering
  \includegraphics[width=\columnwidth]{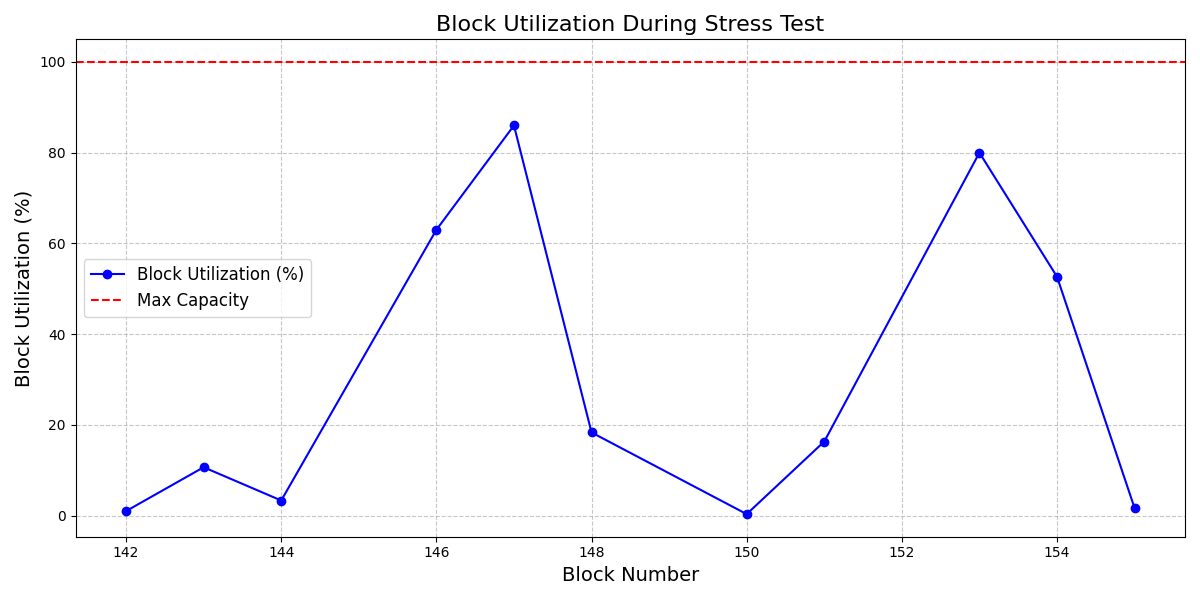}
  \caption{Block utilization during stress test. Early and late phases show inefficiencies due to sequencer startup latency and batch finalization. Peak efficiency achieved mid-test.}
  \label{fig:blockutilization}
\end{figure}

Figure~\ref{fig:blockutilization} illustrates the block utilization rate throughout the stress test, which reflects how efficiently each block's capacity was used. We define 100\% utilization as 300 swap transactions per block—an upper bound derived from the optimized memory footprint of such transactions.

During the initial phase (blocks 142 to 144), utilization was notably low. This corresponds to the transactions sent in the first 1–2 seconds, as shown in Figure~\ref{fig:transactiongenerator}. The underutilization here can be attributed to the sequencer's initialization phase. Although the sequencer was live prior to the test, it had been idle and needed time to warm up, process the initial backlog, and begin forming batches. These early blocks act as a cold start buffer, and while they appear inefficient, they represent necessary system bootstrapping overhead.

Following this, from block 145 onward, utilization quickly rises and stabilizes near full capacity. This improvement coincides with the sequencer processing a steady influx of transactions (3–6 seconds into the test). However, at block 150, utilization sharply drops again—suggesting a shift in sequencer priorities. At this point, the sequencer likely began focusing on finalizing a batch and preparing its submission to Ethereum L1, temporarily deprioritizing transaction inclusion.

A similar pattern appears after block 153, where utilization drops further. By this stage, the backlog had largely been cleared, and the number of new transactions diminished. Additionally, the sequencer again likely prioritized finalization and L1 submission, causing fewer transactions to be added to blocks.

\subsubsection*{\textbf{Key Finding}: Trade-offs in Sequencer Behavior}
These observations highlight a key trade-off in ZK rollup systems: the balance between throughput and timely batch finalization. In our setup, the sequencer appears to switch roles too aggressively—rapidly toggling between transaction processing and batch finalization—which leads to suboptimal resource use. Future designs may benefit from more adaptive scheduling mechanisms that balance these tasks more smoothly, especially under bursty load conditions.

\section{Sequencer and Batch Processing Behavior}
\label{sec:further_observability}

This section investigates the internal operation of our ZK rollup under typical conditions. We analyze how batches are formed, transactions are included, and delays arise in both L1 and L2 components. By examining these detailed metrics, we aim to better understand the performance bottlenecks and sequencing logic of the system.

\subsection{Layer-1 Batch Lifecycle}

Understanding how L2 transactions progress is essential for evaluating the end-to-end performance and reliability of rollup-based systems like ZKsync. A sequencers creates batches of L2 transactions that are later sent to the underlying L1 blockchian. 

\subsubsection*{Seal}  
Sealing a transaction refers to its finalization as part of a batch on the Ethereum Layer 1 (L1) blockchain. It marks the end of the transaction lifecycle, indicating that the batch has been verified and is permanently recorded on-chain.

\subsubsection*{State Keeper}  
The state keeper is a core component responsible for maintaining the L2 state. In ZK rollup systems such as ZKsync, it manages state transitions based on transactions aggregated and executed off-chain before being finalized on L1.

\subsubsection*{Eth Sender}  
The \texttt{eth\_sender} module handles the submission of finalized L2 batches to the L1 blockchain. It is responsible for pushing proofs, commits, and execution data to Ethereum.

\begin{figure*}[h!]
\centering
  \includegraphics[width=0.8\textwidth]{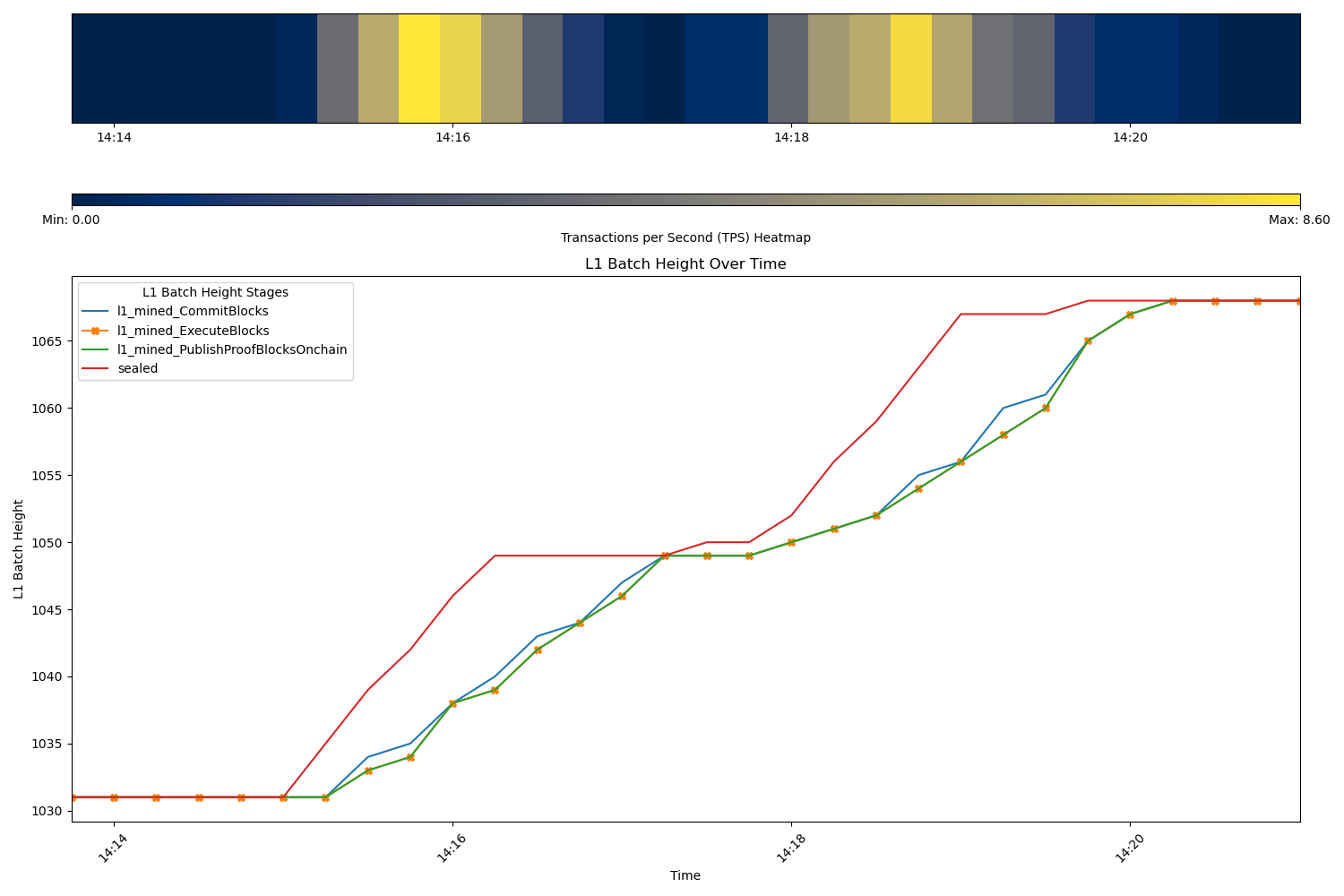}
  \caption{L1 batch height and transaction activity. Shows the progression of batches from sealing to proof publication. Heatmap highlights the correlation between transaction throughput and batch processing stages.}
  \label{fig:Heatmap_L1_batch}
\end{figure*}

Figure~\ref{fig:Heatmap_L1_batch} presents a heatmap correlating L1 batch height with two moderate transaction peaks. Batch stages—such as sealing, proof submission, and execution—are mapped along the timeline, while color intensity indicates transaction throughput (TPS). In this observation window, TPS peaks at 8.6 and drops toward zero at lower activity intervals, at which point the distinctions between batch stages become less visible.

The overall batch height is relatively high due to prior network activity, and this snapshot was chosen to examine how batch stage transitions behave under varying transaction loads. Specifically, it traces batch progression from sealing in the state keeper to final execution by Sepolia validators. Notably, the \texttt{l1\_mined\_ExecuteBlocks} event denotes the point at which a batch is formally included in a mined L1 block.

Delays in the L1 proof stage (\texttt{l1\_mined\_PublishProofBlocksOnChain}) often suggest issues with the prover infrastructure. In contrast, delays in the commit phase (\texttt{l1\_mined\_CommitBlocks}) may indicate bottlenecks or malfunctions in the \texttt{eth\_sender} component.

\subsubsection*{\textbf{Key Finding}: L1 Batch Processing Sensitivity to System Bottlenecks}
The efficiency of L1 batch processing is highly sensitive to network activity levels. Under moderate load, clear distinctions between batch stages are visible, enabling better monitoring of system health. However, delays in L1 proof and commit operations—often due to prover and submission bottlenecks—highlight critical dependencies in the rollup architecture that must be robustly managed for scalability.

For a more detailed batching metrics, see Appendix~\ref{sec:appendix_sequencer_insights}.

\section{Discussion}
\label{sec:discussion}

Our findings indicate that ZK rollups, particularly ZKsync, offer substantial scalability benefits for DeFi applications, yet they introduce nuanced design trade-offs that impact reliability, performance, and operational complexity. This section contextualizes our results, reflecting on the practical implications of the implementation, the limitations observed during testing, and the deeper structural considerations for protocol builders.

\subsection{Deployment and Operational Feasibility}

Deploying and managing a local ZKsync chain with a forked Uniswap V2 exchange exposed the infrastructure demands and limitations of current rollup technology. Although our setup achieved a peak throughput of 71 TPS—well above Ethereum L1’s 15--23 TPS—this required a high-performance server environment (16-core CPU, 64 GB RAM, NVIDIA T4 GPU). Stress testing further revealed that running more than five concurrent transaction generators caused sequencer instability, resulting in failed block production. This underscores a critical operational fragility: even under controlled conditions, rollup components such as the sequencer are susceptible to overload, functioning effectively as single points of failure.

\subsection{Sequencer Scheduling and Bottlenecks}

Observability into the sequencer’s runtime behavior highlighted inefficient toggling between transaction ingestion and batch finalization. In bursty conditions, the sequencer aggressively switched roles, leading to underutilization of resources. This behavior caused significant throughput degradation beyond a modest parallel load. Furthermore, we found that miniblocks frequently included far fewer transactions than the corresponding L1 batches. This gap illustrates the tension between maintaining low latency for user experience and optimizing batch size for L1 cost efficiency.

\subsubsection*{\textbf{Key Finding}: Trade-off in Sequencer Scheduling}

Sequencer responsiveness comes at the cost of batching efficiency. Systems prioritizing soft-finality latency often include miniblocks with suboptimal size, reducing amortization of fixed L1 submission costs.

\subsection{Latency and Finality Guarantees}

Our transaction inclusion analysis confirms that ZKsync can offer a responsive L2 experience: over 50\% of transactions were included in miniblocks within 2.5 seconds. However, this soft finality is not synonymous with settlement. The average time to hard finality—i.e., when batches are verified and finalized on L1—ranged from 10 to 20 minutes, depending on network congestion and prover availability. 

\subsubsection*{\textbf{Key Finding}: Latency Bifurcation}

ZK rollups offer rapid soft finality but delayed settlement. While suitable for most DeFi use cases, this latency bifurcation should be accounted for in systems with settlement-critical logic, such as liquidations or cross-chain bridges.

\subsection{L1 Batch Sealing Insights}

In-depth timing analysis of batch sealing stages revealed that Merkle tree computation is the dominant bottleneck, accounting for up to 2.44 seconds per batch. Other stages—like batch header insertion, log deduplication, or initial writes—completed in under 250 milliseconds. Despite this, total sealing time appeared lower than the sum of all stages, highlighting the role of parallelization.

\subsubsection*{\textbf{Key Finding}: Merkle Tree Updates as a Bottleneck}

Merkle tree updates dominate the L1 sealing process and constrain scalability. Future optimizations in prover infrastructure and cryptographic batching may be required to mitigate this bottleneck.

\subsection{Efficiency Disparities Between Miniblocks and L1 Batches}

Figure~\ref{fig:Txs_miniblock_L1} highlighted that miniblocks typically contain over 3$\times$ fewer transactions than L1 batches. This gap arises because miniblocks prioritize responsiveness and proof readiness over throughput. While this behavior enhances user experience, it results in suboptimal gas amortization.

\subsubsection*{\textbf{Key Finding}: Batching Trade-offs Reduce Efficiency}

Miniblocks enable fast execution but at the cost of batching efficiency. Aligning miniblock size more closely with L1 batch capacity could improve cost-effectiveness in production settings.

\subsection{Code Migration and Compatibility Challenges}

Forking and adapting Uniswap V2 to run on ZKsync, though successful, required substantive engineering effort. Differences in compiler versions (e.g., transition to Solidity 0.8.x), stricter type handling, and error propagation models revealed non-trivial incompatibilities. These challenges caution against assumptions of seamless EVM compatibility in rollup environments.

\subsubsection*{\textbf{Key Finding}: EVM Compatibility is Nuanced}

While rollups aim for EVM equivalence, migration of complex contracts reveals subtle but critical divergences. Protocol teams should budget for code audits, formal verification, and behavior validation during migration.

\subsection{Summary: Practical Takeaways}

Our implementation and stress testing of a live ZKsync rollup instance connected to a forked Uniswap V2 exchange revealed both the performance promise and architectural limitations of current ZK rollup infrastructure. These takeaways offer guidance for researchers, developers, and infrastructure teams.

\begin{itemize}
    \item \textbf{Scalability:} Our testbed achieved 71 TPS with gas-intensive swaps—over 3$\times$ higher than Ethereum L1.
    \item \textbf{Latency:} Over 50\% of swaps were confirmed in under 2.5s, enabling near-instant UX.
    \item \textbf{Sequencer Bottlenecks:} Current designs are fragile under load; stability mechanisms and decentralization are urgent needs.
    \item \textbf{Merkle Tree Performance:} Tree updates are the primary source of latency in L1 batch sealing.
    \item \textbf{Efficiency Gap:} L1 batches amortize costs better than miniblocks, which remain underfilled.
    \item \textbf{Code Porting Overhead:} Rollup EVM compatibility simplifies migration, but edge-case divergences demand careful engineering.
    \item \textbf{Operational Cost:} Self-hosting a rollup node requires significant computational resources, which may exclude smaller teams or individual developers.
\end{itemize}

\section{Conclusion}
\label{sec:conclusions}

This paper presents the design and evaluation of a real-time ZK rollup proof-of-concept based on ZKsync, integrated with a forked Uniswap V2 exchange. Our empirical stress tests demonstrated that the system can reach 71 TPS of gas-intensive swap transactions with soft finality under 2.5 seconds, while supporting observability and security instrumentation in a fully local environment.

We identified centralization of the sequencer and prover as primary constraints on scalability and fault tolerance, proposing container-based replication as a practical solution. Moreover, we detailed the risks and engineering challenges associated with migrating DeFi protocols to rollups, especially under partial EVM compatibility.

ZK rollups are a critical step forward in Ethereum's scalability roadmap. Yet realizing their full promise requires further innovation in decentralization, fault tolerance, and developer tooling. Our platform and results serve as a blueprint for future research and production-grade deployments of high-throughput DeFi applications on Layer 2.

\bibliographystyle{IEEEtran}
\bibliography{main}

\appendix
\section*{Appendix: Detailed Batch Sealing Breakdown}
\label{sec:appendix_sequencer_insights}

This appendix provides a deeper view into the internal execution stages involved in finalizing Layer-1 batches within the ZK Rollup system. These figures complement the main discussion in Section~\ref{sec:further_observability} by offering stage-level granularity on performance metrics. 

\subsection{Sequencer Latency}

\begin{figure*}[h!]
\centering
  \includegraphics[width=0.75\textwidth]{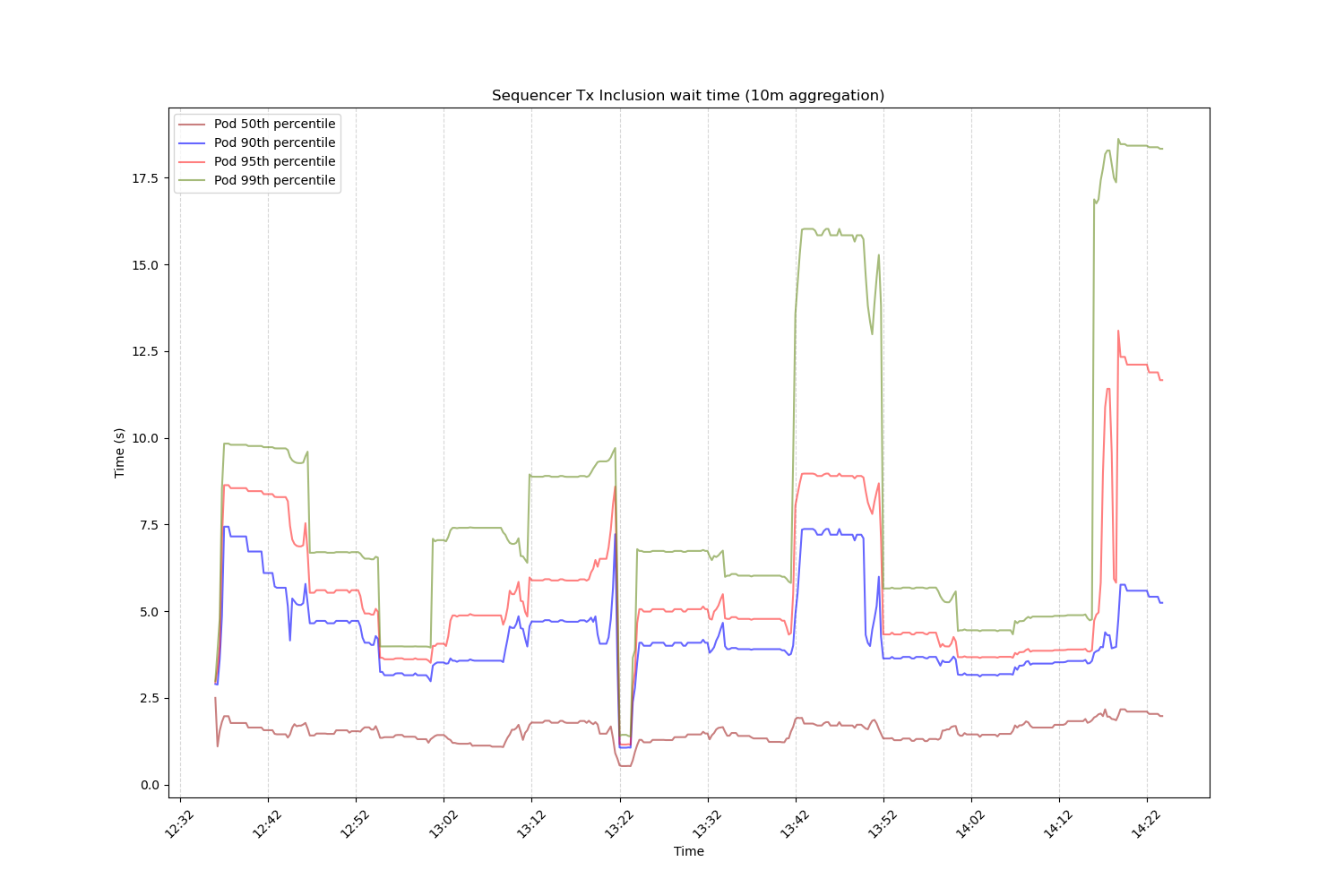}
  \caption{Latency distribution between transaction receipt and miniblock inclusion. 50\% of transactions are included within 2.5s, showing responsiveness of sequencer in moderate load conditions.}
  \label{fig:Tx_inclusion_wait}
\end{figure*}

Figure~\ref{fig:Tx_inclusion_wait} illustrates how long transactions spend in the L2 mempool before being included in a miniblock. The median latency is around 2.5 seconds. This metric excludes other contributing factors to user-perceived delay such as gas estimation and polling intervals. Polling time is the interval during which a client or application checks the status of a transaction in the network.

\subsection{Miniblock vs. L1 Batch Efficiency}
\textit{Miniblock.}
A miniblock refers to a smaller set of transactions that are batched together for processing before being included in a larger block. In ZKsync, miniblocks optimize transaction throughput by grouping multiple transactions, thus reducing latency and increasing efficiency.

\begin{figure*}[h!]
\centering
  \includegraphics[width=0.9\textwidth]{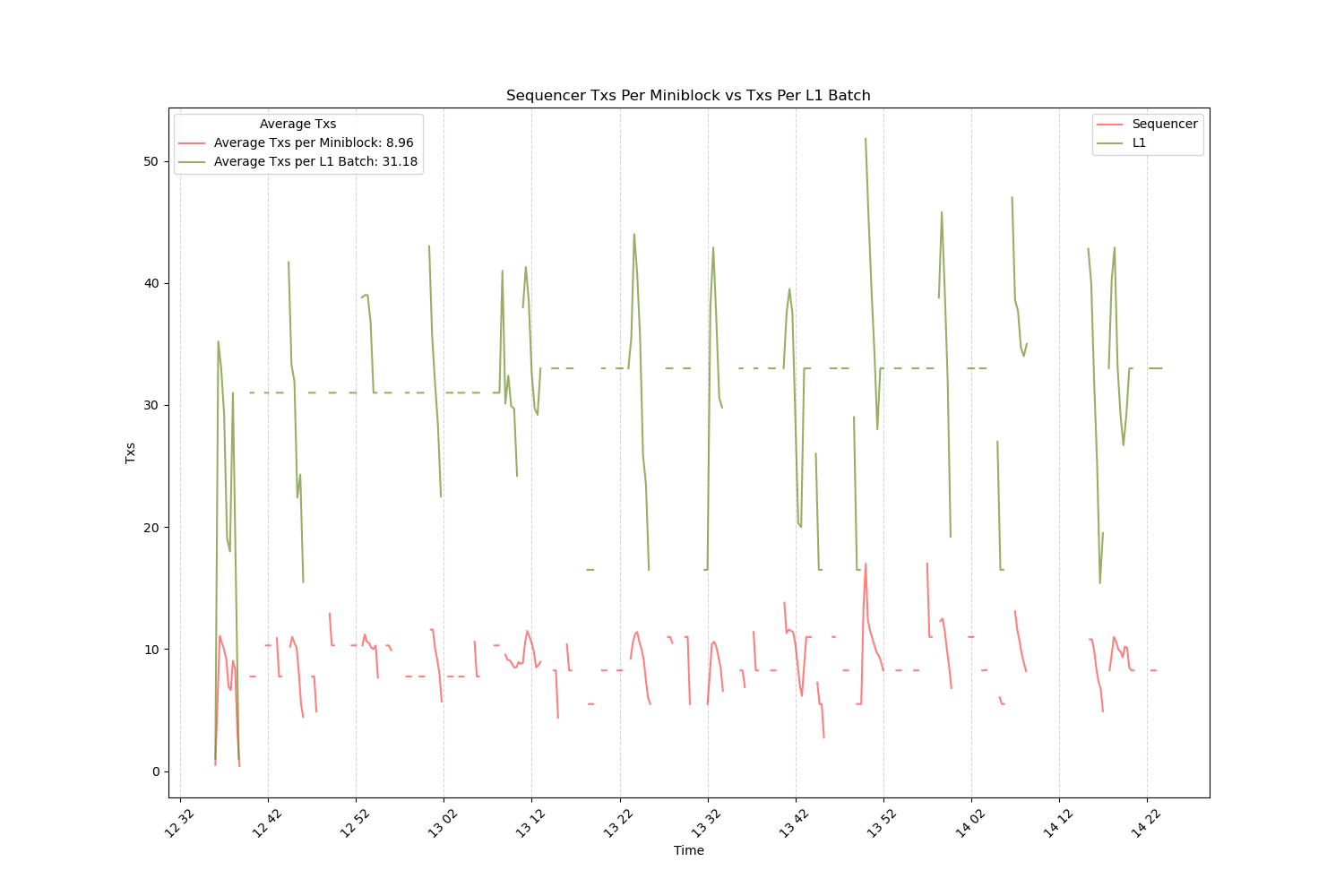}
  \caption{Transactions per miniblock vs. L1 batch. Illustrates efficiency gains through batching: L1 batches include over 3\texttimes{} more transactions than individual miniblocks.}
  \label{fig:Txs_miniblock_L1}
\end{figure*}

Figure~\ref{fig:Txs_miniblock_L1} compares the number of transactions per miniblock in the ZKsync sequencer with those per L1 batch. L1 batches typically include significantly more transactions, as they are optimized to maximize block space and throughput. In contrast, miniblocks in ZKsync tend to include fewer transactions, reflecting their dual role: ensuring low-latency processing on L2 and preparing transactions for computationally intensive rollup and proof generation. This design highlights a key trade-off in L2 scaling between responsiveness and batching efficiency.

\subsubsection*{\textbf{Key Finding}: Miniblock Responsiveness vs. L1 Batching Efficiency}
L2 systems like ZKsync achieve faster responsiveness via smaller, more frequent miniblocks. However, this comes at the cost of batching efficiency—L1 batches consolidate over three times more transactions, maximizing throughput and minimizing on-chain costs. This underscores a fundamental trade-off between low latency and block space optimization in rollup architectures.

\subsection{L1 Batch Sealing Time}

\textit{Commit L1 Batch.}  
In the context of L2 systems, this refers to the process of finalizing and recording a batch of L2 transactions onto the Ethereum mainchain. It marks the transition from soft to hard finality.

\textit{Fictive Miniblock.}  
A fictive miniblock is a simulated construct used internally for system operations such as testing, optimization, or performance tuning. Unlike regular miniblocks, these are not broadcast to the blockchain but assist in preparing the system for real transactions.

\textit{Filter Written Slots.}  
This step filters and organizes the transaction data prior to block inclusion, resolving conflicts and ensuring data integrity before the write phase.

\textit{Insert Initial Writes.}  
Represents the stage where preliminary transaction data is written to internal databases, initiating the batch formation process.

\textit{Insert L1 Batch Header.}  
The process of appending a metadata header to a transaction batch, which prepares the data for submission and commitment to the L1 blockchain.

\textit{Log Deduplication.}  
A stage where duplicate log entries are identified and removed, improving storage efficiency and reducing unnecessary computation.

\textit{Set L1 Batch Number for Miniblocks.}  
Assigns an L1 batch identifier to each miniblock, ensuring correct alignment between L2 sequencing and L1 batch submission.

\textit{Waiting for Tree.}  
This phase involves waiting for updates to the Merkle tree, a core cryptographic structure ensuring the integrity and verifiability of state transitions. This step is computationally intensive and a common performance bottleneck.

\begin{figure*}[h!]
\centering
  \includegraphics[width=0.9\textwidth]{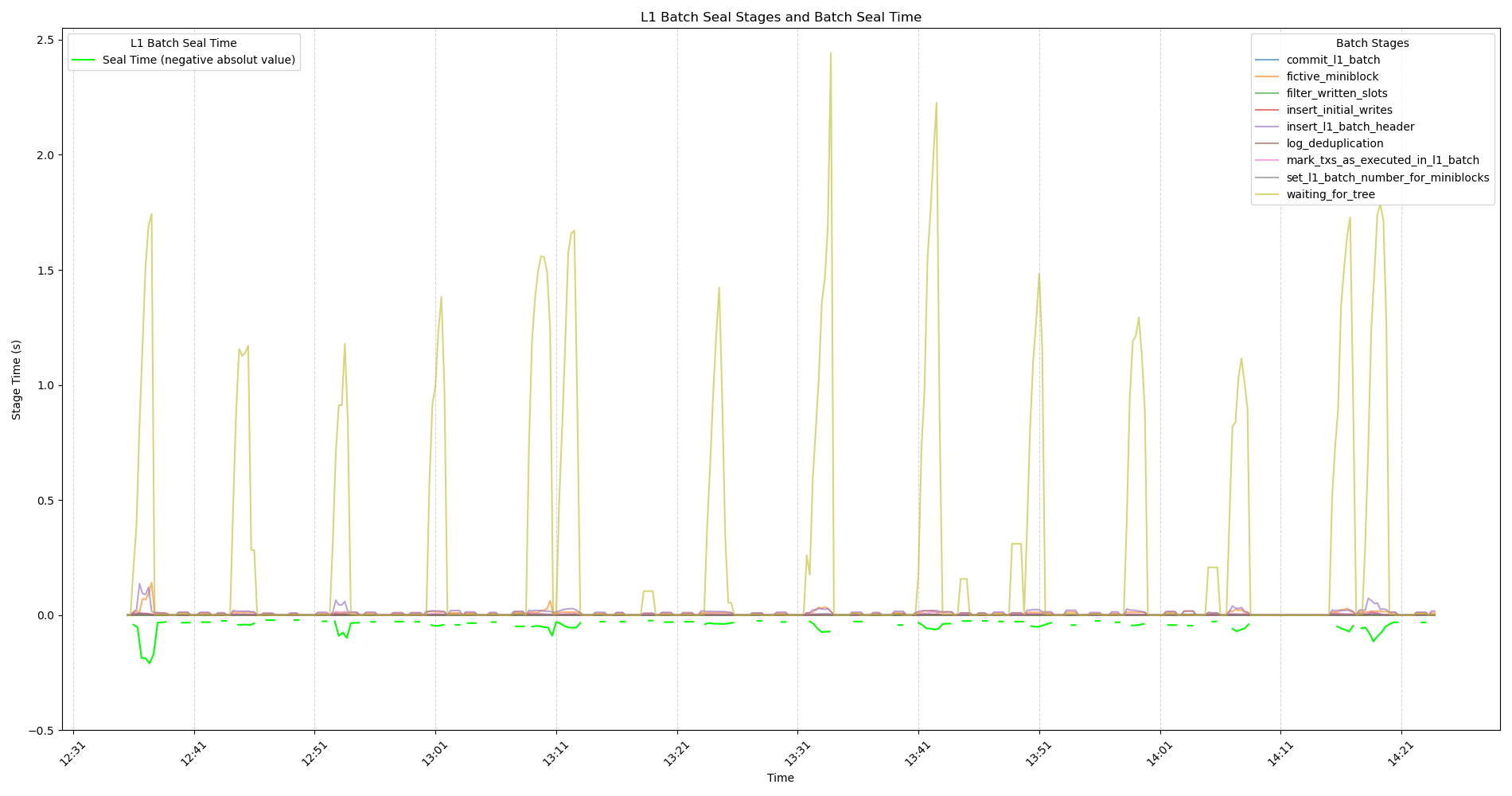}
  \caption{Processing time of L1 batch sealing stages. Highlights Merkle tree update as a performance bottleneck with a 2.44s delay. Other stages complete within 250ms.}
  \label{fig:L1_stages}
\end{figure*}

Figure~\ref{fig:L1_stages} illustrates the time spent in each L1 sealing stage. Most components, such as batch header insertion, log deduplication, and initial writes, complete in under 250 milliseconds. In contrast, the “Waiting for Tree” phase dominates the timeline with a peak delay of 2.44 seconds—making it the primary bottleneck in the sealing process.

Notably, the total L1 batch sealing time may appear shorter than the sum of individual stage durations. This discrepancy arises due to parallel execution: many operations are pipelined or overlap in time. Such concurrent processing is a hallmark of optimized distributed systems and critical to maintaining high throughput.

\subsubsection*{\textbf{Key Finding}: Merkle Tree Updates as a Bottleneck}
While most batch sealing stages complete in under 250ms, the Merkle tree update stage introduces a significant delay—up to 2.44s—making it the dominant performance bottleneck. Optimizing this step is crucial for improving batch finalization times and sustaining L2 throughput at scale.

\subsection{Detailed Batch Sealing Breakdown}

This section provides a deeper view into the internal execution stages involved in finalizing Layer-1 batches within the ZK Rollup system. 

\begin{figure*}[h!]
\centering
  \includegraphics[width=1\textwidth]{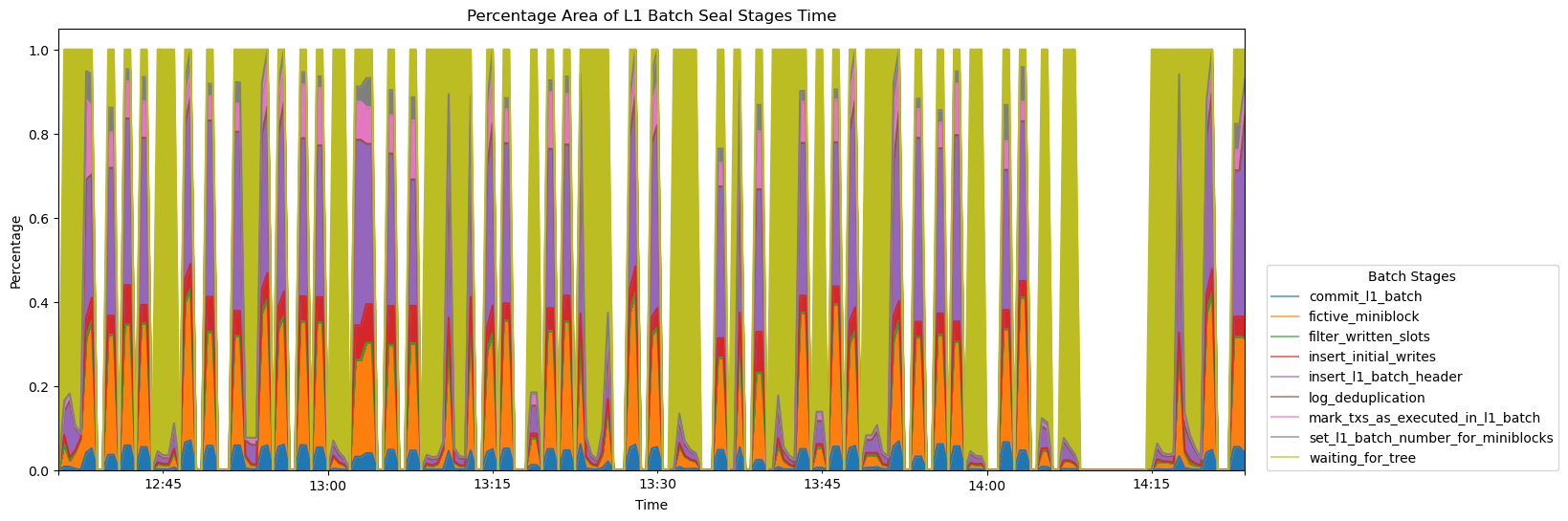}
  \caption{Percentage share of each stage in total L1 batch sealing time. Confirms the dominant contribution of Merkle tree computation and illustrates benefits of parallelization.}
  \label{fig:L1_stages_percentage}
\end{figure*}

Figure~\ref{fig:L1_stages_percentage} illustrates the relative contribution of each sealing stage to the overall processing time. It confirms the Merkle tree update as the primary bottleneck, suggesting that further performance gains could be achieved by optimizing this computation or enabling concurrency in surrounding steps.

\begin{figure*}[h!]
\centering
  \includegraphics[width=1\textwidth]{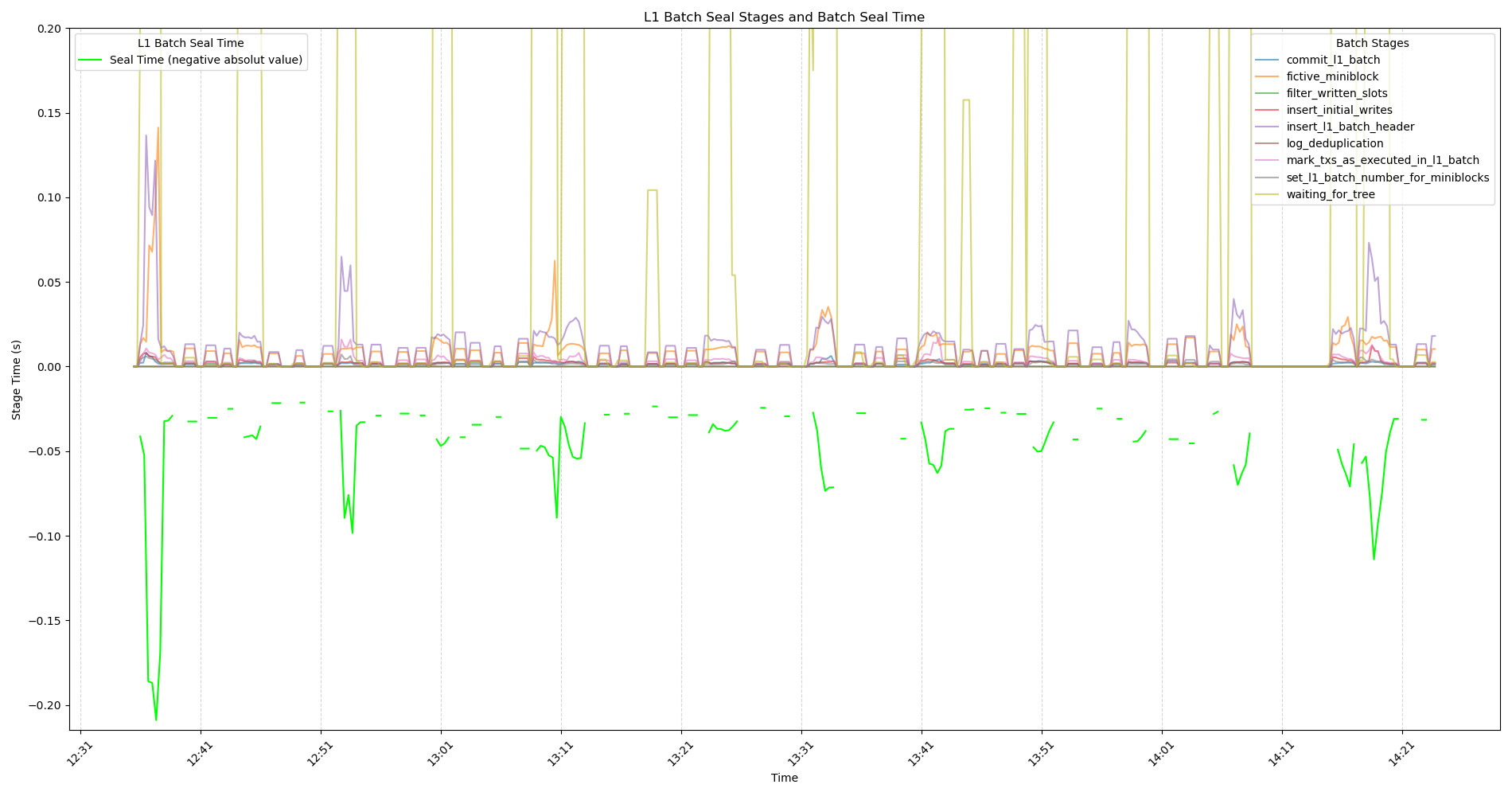}
  \caption{Detailed breakdown of L1 sealing stage durations. Emphasizes stage-level granularity for optimization in future prover implementations.}
  \label{fig:L1_stages_detail}
\end{figure*}

Figure~\ref{fig:L1_stages_detail} provides an absolute timing breakdown of individual sealing components. This detailed view is useful for implementers seeking to fine-tune specific components of the prover pipeline and reveals the extent

\end{document}